\documentclass[prb,twocolumn,showpacs,preprintnumbers,amsmath,amssymb,floats]{revtex4-1}
\usepackage{graphicx}
\usepackage{amsmath}
\usepackage{epstopdf}
\usepackage{amsbsy}
\usepackage{color}
\usepackage{lipsum}
\usepackage{dcolumn}% Align table columns on decimal point
\usepackage{bm}% bold math

\begin{document}
\title{Competing magnetic double-Q phases and superconductivity-induced re-entrance of $C_2$ magnetic stripe order in iron pnictides}
\author{Maria N. Gastiasoro and Brian M. Andersen}
\affiliation{Niels Bohr Institute, University of Copenhagen, Universitetsparken 5, DK-2100 Copenhagen,
Denmark}
\date{\today}

\begin{abstract}
% ------------------------------------------------------------------------------------------------------------------------------------------------------ %
We perform a microscopic theoretical study of the generic properties of competing magnetic phases in iron pnictides. As a function of electron filling and temperature, the magnetic stripe (single-Q) order forms a dome, but competing non-collinear and non-uniform double-Q phases exist at the foot of the dome in agreement with recent experiments. We compute and compare the electronic properties of the different magnetic phases, investigate the role of competing superconductivity, and show how disorder may stabilize double-Q order. Superconductivity is shown to compete more strongly with double-Q magnetic phases, which can lead to re-entrance of the $C_2$ (single-Q) order in agreement with recent thermal expansion measurements on K-doped Ba-122 crystals.  
\end{abstract}
% ===================================================================================== %
\maketitle

In correlated materials in general, and unconventional superconductors in particular, a microscopic understanding of the magnetism is of paramount importance. Generally, this is because a proper description of the relevant exchange mechanism in these materials is intimately tied to their basic electronic properties. More specifically, it is additionally shown within a wide class of models that the nature of the magnetic fluctuations may be closely linked to the emergence of the superconducting condensate.\cite{chubukovreview,scalapinoreview,astridPM}

\begin{figure}[t]
\begin{center}
\includegraphics[width=0.8\columnwidth]{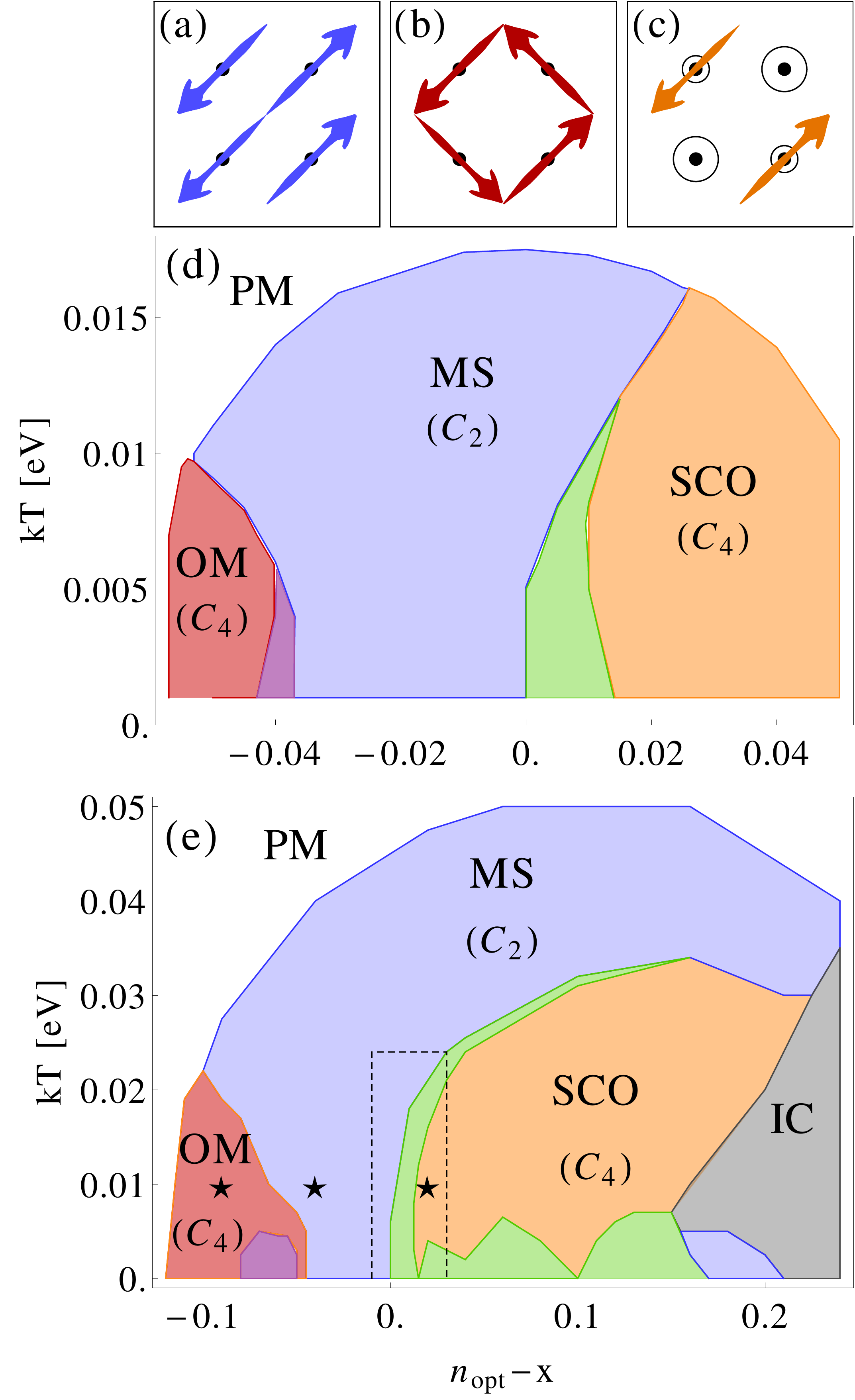}
\end{center}
\caption{(Color online) Spin and charge order of the (a) 1Q MS ($\mathbf{M_2}=0$), (b) 2Q OM ($\mathbf{M_1}\perp \mathbf{M_2}$), and (c) SCO ($\mathbf{M_1}=\mathbf{M_2}$). The black circles in (c) represent the $\mathbf{Q_1}+\mathbf{Q_2}=(\pi,\pi)$ charge order.
(d,e) Magnetic phase diagrams as obtained from Eq.~(\ref{eq:H}) as a function of $T$ and filling $n_{opt}-x$ for (d) $U=0.85$ eV and (e) $U=0.95$ eV. 
The green (purple) area indicates regions of coexisting MS and SCO (OM).}
\label{fig:1}
\end{figure}

Focussing on the iron-based superconductors, the prevalent magnetic structure consists of collinear magnetic stripe (MS) order with in-plane moments oriented antiferromagnetic (ferromagnetic) along the $a$ ($b$) axis of the orthorhombic Fe lattice as shown in Fig.~\ref{fig:1}(a). Thus, this configuration of moments singles out the $\mathbf{Q}_1=(\pi,0)$ ordering vector (1Q), i.e. $\mathbf{M}(r)=\mathbf{M}_1 \exp(i \mathbf{Q}_1 \cdot \mathbf{r})$. An obvious question, however, is why the system does not take advantage of the enhanced susceptibility at both $\mathbf{Q}_1$ and $\mathbf{Q}_2=(0,\pi)$ to form other magnetic phases, e.g. double-Q (2Q) phases consisting of superpositions of ordering at $\mathbf{Q}_1$ and $\mathbf{Q}_2$ with $\mathbf{M}(r)=\sum_{l=1,2} \mathbf{M}_l \exp(i \mathbf{Q}_l \cdot \mathbf{r})$. This question has been studied theoretically mainly using various effective field theories restricted to the vicinity of the magnetic transition temperature $T_N$.\cite{lorenzana08,eremin2010,brydon11,giovannetti11} These works have identified two competing magnetic structures of the 2Q type: 1) an orthomagnetic (OM) non-collinear phase with nearest neighbor moments at right angles as shown in Fig.~\ref{fig:1}(b), and 2) a collinear non-uniform spin and charge ordered (SCO) phase as shown in Fig.~\ref{fig:1}(c). The favorable magnetic order depends delicately on the band structure, doping level, and interactions.\cite{lorenzana08,eremin2010,brydon11,giovannetti11} 

Experimentally, the dominating magnetic order in the iron pnictides is the 1Q MS state. This phase lowers the $C_4$ symmetry of the high-$T$ tetragonal phase to orthorhombic $C_2$, and causes an associated splitting of the crystal Bragg peaks due to magneto-elastic coupling. Recently, several experiments have, however, reported the discovery of magnetic order without an associated structural splitting, i.e. in the tetragonal phase,\cite{kim10,avci14} which has been taken as indirect evidence for a magnetic driven structural transition in the case of 1Q MS order.\cite{wangcondmat14} In the case of Ba(Fe$_{1-x}$Mn$_x$)$_2$As$_2$,\cite{kim10} however, additional experiments have shown that Mn induce local regions of magnetic $(\pi,\pi)$ order and a crystal structure consistent with intertwined short-range clusters of both tetragonal and orthorhombic structure.\cite{tucker12,texier12,leboeuf13,inosov13} The collective outcomes of these  experiments have recently been shown to be captured within a microscopic disorder scenario.\cite{gastiasoro14} 

The study of 2Q order in Ba(Fe$_{1-x}$Mn$_x$)$_2$As$_2$ should be contrasted to other pnictides with out-of-plane dopants where a disorder scenario seems less relevant. This includes for example Ba-122 doped with Na or K where experiments have found evidence for a phase transition into a long-range ordered magnetic phase with tetragonal crystal structure.\cite{avci14,hassinger12,boehmer14,allred1,mallett1,allred2,mallett2} For the case of hole doped Ba$_{1-x}$Na$_x$Fe$_2$As$_2$ this magnetic phase exists at the foot of the magnetic dome in the phase diagram.\cite{avci14} More recently, B\"{o}hmer {\it et al.}, used thermal expansion measurements to find a tetragonal low-$T$ phase consistent with a magnetic 2Q phase, and additionally revealed a superconductivity-induced re-entrant orthorhombic phase at even lower $T$.\cite{boehmer14} Collectively, these experimental findings define the following main challenges for a theoretical description: 1) the existence of 2Q phases in restricted (intermediate) doping regimes limited by MS order and superconductivity (SC), 2) The 2Q phases exist in a limited (intermediate) $T$ range, 3) SC competes with magnetic order causing a lowering of $T_c$ upon entering the magnetic 2Q phase from the paramagnet, and 4) SC competes more strongly with 2Q phases than the $C_2$ MS as seen by an upward $T_c$ jump when transitioning from the coexistence phase of SC and 2Q magnetism to a coexistence phase of MS order and SC.\cite{boehmer14}

Here we perform a study of the stability of, and phase transitions between, the competing magnetic phases obtained from a five-band Hamiltonian relevant to the iron pnictides. The Coulomb interaction is treated within unrestricted self-consistent Hartree-Fock, i.e. all charge and spin densities are allowed to vary at each separate site. These calculations constitute a comprehensive microscopic study of the 2Q magnetic phases, also in the presence of SC and disorder, and allows to access the entire $T$ regime contrary to previous theoretical studies.\cite{lorenzana08,eremin2010,brydon11,kang14}  The model provides an explanation of all four challenges outlined above.  

The starting Hamiltonian consists of a five-orbital tight-binding band relevant to the pnictides~\cite{ikeda10} and the usual multiband Hubbard-Hund interaction term. When mean-field decoupled  this leads to the following model

% ------------------------------------------------------------------------------------------------------------------------------------------------------ %

\begin{eqnarray}
\mathcal{H}^{MF}=\sum_{ij\mu\nu}
\begin{pmatrix}
 \hat{c}_{i\mu\uparrow}^{\dagger} & \hat{c}_{i\mu\downarrow}^{\dagger}
\end{pmatrix}
\begin{pmatrix}
\varphi_{ij\uparrow}^{\mu\nu} & \omega_{ii\uparrow}^{\mu\nu}  \\
\omega_{ii\downarrow}^{\mu\nu} & \varphi_{ij\downarrow}^{\mu\nu}  
\end{pmatrix}
\begin{pmatrix}
 \hat{c}_{j\nu\uparrow}\\\hat{c}_{j\nu\downarrow}
\end{pmatrix},
\label{eq:H}
\end{eqnarray}
where $c_{i \mu\sigma}^{\dagger}$ creates an electron at site $i$ with spin $\sigma$ in orbital state $\mu$.
$\varphi_{ij\sigma}^{\mu\nu}$ and $\omega_{ii\sigma}^{\mu\nu}$ are functions of the interaction parameters $U$, $J$ ($J=U/4$), and the fields $\langle \hat{c}_{i\mu\sigma}^{\dagger} \hat{c}_{j\nu\sigma'}\rangle$ (see Supplementary Information (SI)). We diagonalize Eq.\eqref{eq:H} unrestricted on $N_x \times N_y$ lattices, and self-consistently calculate the spin  $M^l(\mathbf{r_i})=\sum_{\mu\sigma\sigma'} \langle \hat{c}_{i\mu\sigma}^{\dagger}  \tau_{\sigma\sigma'}^l \hat{c}_{i\mu\sigma'}\rangle$, and charge density $n(\mathbf{r_i})=\sum_{\mu} \left(\langle \hat{c}_{i\mu\uparrow}^{\dagger} \hat{c}_{i\mu\uparrow}\rangle+\langle \hat{c}_{i\mu\downarrow}^{\dagger} \hat{c}_{i\mu\downarrow}\rangle\right)$,
where $l=x,z$, and extract the ordering components $\mathbf{M_l}$ and $n_l$ of $\mathbf{M}(\mathbf{r})=\sum_{l} \mathbf{M}_l \exp(i \mathbf{Q}_l \cdot \mathbf{r})$ and  $n(\mathbf{r})=\sum_{l} n_l \exp(i \mathbf{Q}_l \cdot \mathbf{r})$. 
We have compared the free energy of the magnetic states $\mathcal{F}=\langle\mathcal{H}^{MF}\rangle-T\mathcal{S}$ to verify the stability of the results.

Figure~\ref{fig:1}(d,e) display two representative the phase diagrams as obtained from self-consistently diagonalizing Eq.~(\ref{eq:H}) as a function of $T$ and electron filling for  $U=0.85$ eV (d) and $U=0.95$ eV (e). 
$n_{opt}=5.91$ is defined as the electron filling with the highest $T_N$ as deduced by the paramagnetic susceptibility, and the total filling is $\langle n \rangle = n_{opt} -x$. The fact that the optimal doping level for the magnetic order is offset from $n=6$ for DFT-generated bands is well known,\cite{schmiedt12} and not important for the conclusions of this paper.
As seen from both cases, the 2Q phases, OM and SCO, exist at the foot of the MS dome on the electron and hole-doped side respectively, and whereas the transition between the MS and OM phases is sharp, a more gradual transition takes place between the MS and the SCO phases as indicated by the green intermediate regions. 
Interestingly, both recent M\"{o}ssbauer spectroscopy studies of Sr$_{1-x}$Na$_x$Fe$_2$As$_2$~\cite{allred2} and moun spin rotation measurements on Ba$_{1-x}$K$_x$Fe$_2$As$_2$~\cite{mallett2} found that indeed the 2Q magnetic phase of the hole-doped system seems to be the SCO phase, in agreement with the phase diagrams in Fig.~\ref{fig:1}(d,e). 
The colinear spin structure of the SCO phase was also recently verified by spin polarized neutrons, additionally finding that the moments are oriented along the $c$-axis.~\cite{wasser15}
The grey area denoted IC in Fig.~\ref{fig:1}(e) represents an incommensurate magnetic phase where the ordering vectors $\mathbf{Q_1}/\mathbf{Q_2}$ no longer faithfully represent the magnetic ground state of the system. The IC phase is absent in Fig.~\ref{fig:1}(d) because the lower $U$ leads to a vanishing magnetization at significantly lower doping levels compared to  Fig.~\ref{fig:1}(e).

\begin{figure}[t]
\begin{center}
\includegraphics[width=0.99\columnwidth]{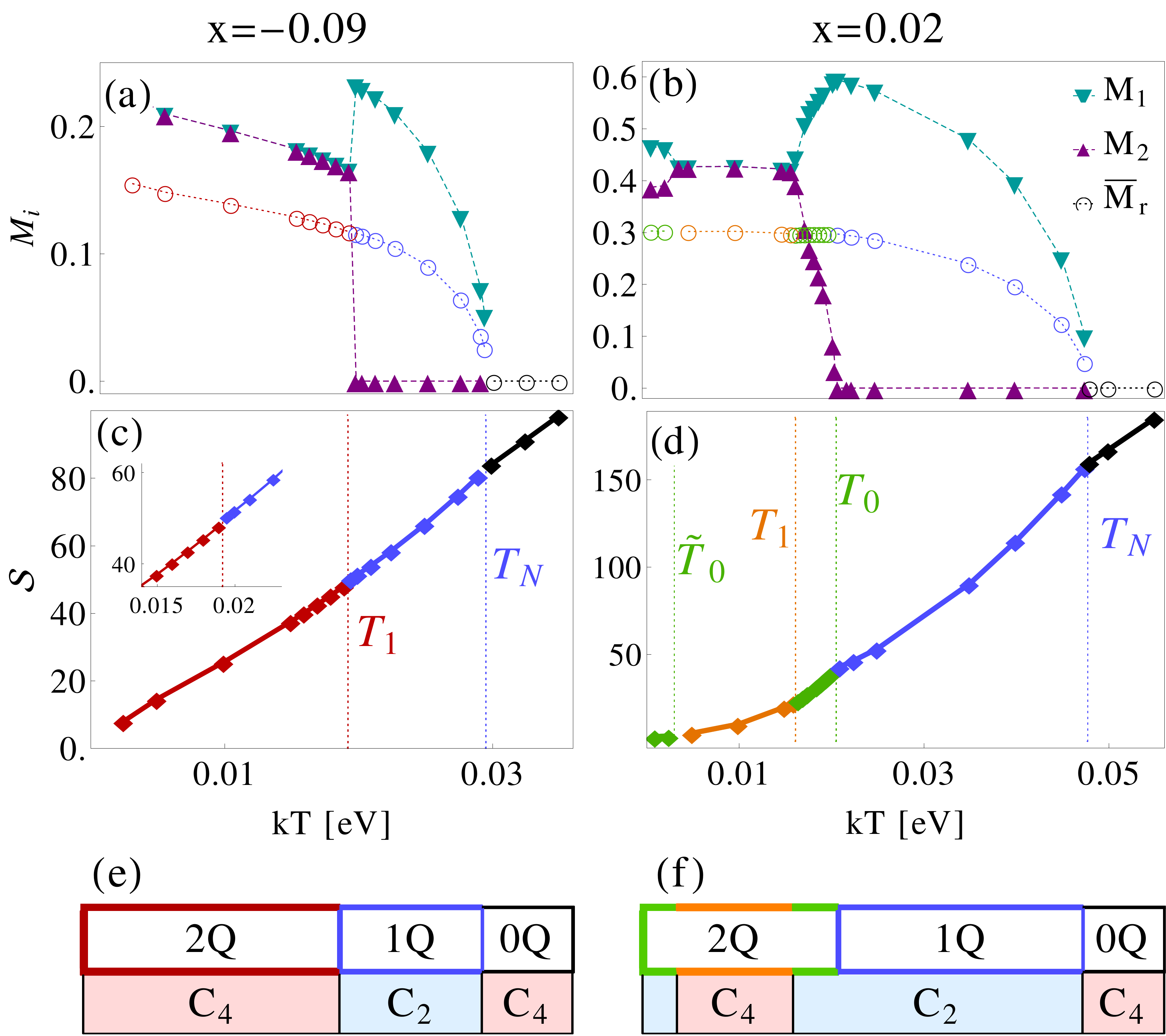}
\end{center}
\caption{(Color online) (a) $T$ evolution at $x=-0.09$ of the magnetic components $M_1$ ($\blacktriangledown$), $M_2$ ($\blacktriangle$) and $\bar M_r=\frac{1}{2} \sqrt{M_1^2+M_2^2}$ ($\circ$), and (c) the entropy $\mathcal{S}(T)$. The color changes in $\bar M_r$ and $\mathcal{S}$ represent the magnetic phase transitions shown also in Fig.~\ref{fig:1}(e). (b,d) The same as (a,c) but for $x=0.02$ with SCO order. (e,f) summarize the $T$-dependence of the magnetic structure and the expected associated lattice symmetry.}
\label{fig:2}
\end{figure}

In the remainder of this paper we focus on the $U=0.95$ eV case, and use the $x=-0.09$ and $x=0.02$ electron and hole fillings, respectively, to discuss the transition from the MS state to the OM and SCO phases upon lowering $T$. Figures~\ref{fig:2}(a,b) show the $T$ dependence of the SDW components $M_1\equiv|\mathbf{M_1}|$ and $M_2\equiv|\mathbf{M_2}|$ for both fillings. At $T_N$, $M_1$ gradually increases while $M_2$ remains zero, signalling a second order transition into the MS state. 
In the $x=-0.09$ case, [Fig.~\ref{fig:2}(a)], at $T_1<T_N$ the 1Q-2Q transition takes place and the moments re-orient to form the OM state with $\mathbf{M_1}\perp \mathbf{M_2}$ and $M_2=M_1$. As seen from Fig.~\ref{fig:2}(a), the sudden jump of $M_2$ at $T_1$ is compensated by a reduction in $M_1$, leaving the average magnetization $\bar M_r=\frac{1}{2} \sqrt{M_1^2+M_2^2}$ nearly unchanged. In Fig.~\ref{fig:2}(c) we display the $T$ dependence of the entropy $\mathcal{S}(T)$. The small discontinuity in $\mathcal{S}$ at $T_1$ [see inset of Fig.~\ref{fig:2}(c)] agrees with a weak first order transition. 

The 1Q-2Q transition taking place at $x=0.02$ is shown in Fig.~\ref{fig:2}(b,d). As seen, in this case the second component $M_2$ continuously increases below $T_0<T_N$ in a second order fashion. The increase of $M_2$ is again compensated by a decrease in $M_1$. In this case, however, $\mathbf{M_2}$ is aligned with $\mathbf{M_1}$, and the spin order remains collinear. As soon as $M_2>0$, a small charge order is induced at $\mathbf{Q_3}\equiv \mathbf{Q_1}+\mathbf{Q_2}=(\pi,\pi)$ which scales with $\mathbf{M_1}\cdot\mathbf{M_2}$ and thus increases gradually (see SI for more information). In the range $T_1<T<T_0$, where $M_2$ is increasing, the system is still $C_2$ symmetric ($0<M_2<M_1$) but has developed characteristics of the SCO state, such as the $(\pi,\pi)$ charge order [see Fig.~\ref{fig:1}(c)]. Finally at $T<T_1$ the pristine $C_4$ symmetric SCO phase with $M_2=M_1$ is formed, but we find that at the lowest $T<\tilde T_0$ a re-entrance to a weakly $C_2$ symmetric 2Q phase occurs ($0<M_2<M_1$). As can be seen from $\mathcal{S}(T)$ in Fig.~\ref{fig:2}(d) weak thermodynamic signatures are expected throughout the $T$ range. The $T$ evolution of both transitions and their associated lattice symmetries  are summarized in Figs.~\ref{fig:2}(e,f). 

\begin{figure}
\begin{center}
\includegraphics[width=0.9\columnwidth]{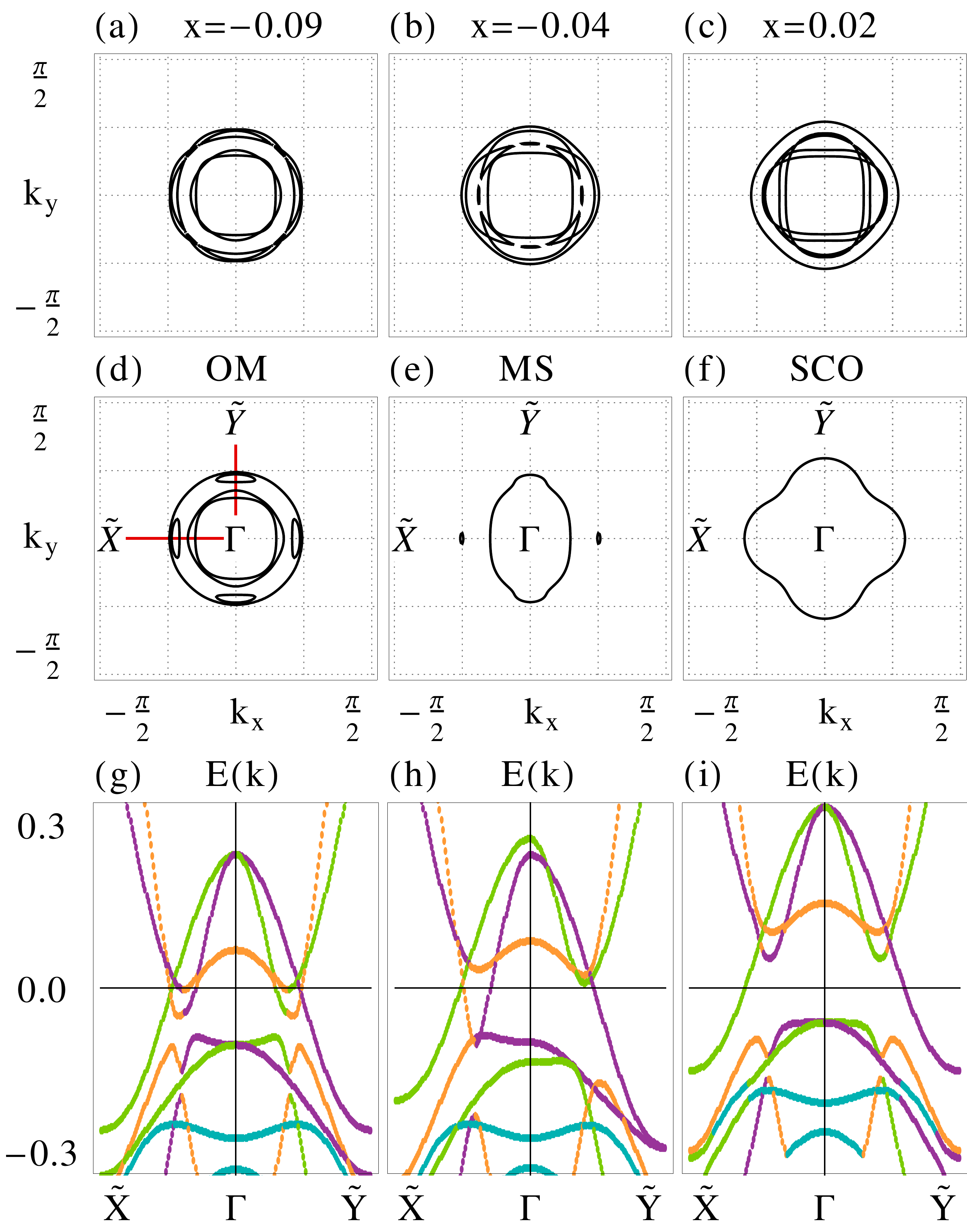}
\end{center}
\caption{Fermi surface in the folded BZ, $-\pi/2<k_x,k_y<\pi/2$, in the PM state for (a) $x=-0.09$, (b) $x=-0.04$ and (c) $x=0.02$, and in the magnetic states (d) OM  ($\mathbf{M_1}=0.2 \hat z$; $\mathbf{M_2}=0.2 \hat x$), (e) MS ($M_1=0.52$; $M_2=0$), and (f) SCO ($M_1=M_2=0.43$). (g-i) Band structure along the momentum path $\tilde X-\Gamma-\tilde Y$ shown in (d) for the (g) OM, (h) MS, and (i) SCO phase. The main orbital contributions are shown by purple: $d_{xz}$; green: $d_{yz}$; orange: $d_{xy}$; cyan: $d_{z^2}$.}
\label{fig:3}
\end{figure}

Next we compare the electronic properties of the magnetic phases, MS, OM, and SCO, by focussing on the three different fillings indicated by the black stars in Fig.~\ref{fig:1}(e).  
In order to illustrate the different nesting conditions, we first show in Figs.~\ref{fig:3}(a)-(c) the FS in the PM state in the folded Brillouin zone (BZ) where the $X$ and $Y$ centered elliptical electron pockets $\beta_1$ and $\beta_2$, and the $M$ centered $\gamma$ hole pocket all fold on top of the $\Gamma$ point (see SI for further details). 
In the ordered state, energy gaps can open at the crossings of the bands connected by the magnetic ordering vector, weighted by the matrix elements $a_{\mu}^n(\mathbf{k})=\langle n|\mu \mathbf{k}\rangle$ which relate orbital and band states. This is apparent in Fig.~\ref{fig:3}(a,d) where the weakly nested $\beta_i$ and the outer hole pocket $\alpha_2$ are gapped around $k_x=\pm k_y$ upon the OM formation. The rest of the SDW gaps are opened below the Fermi energy $\epsilon_F$, and the reconstructed bands are seen from Fig.~\ref{fig:3}(g) to contain considerable orbital mixture. 
As the filling is increased in Fig~\ref{fig:3}(b), multiple electron-hole band crossings get closer to the $\epsilon_F$, and additional nested areas appear at the FS, for example the ones connected by the $\gamma$ and $\beta_i$ pockets. This enhances the spin susceptibility at $\mathbf{Q_1}/\mathbf{Q_2}$, which naturally leads to larger SDW order and more pronounced energy gaps and FS reconstruction in the MS state. The resulting FS in Fig~\ref{fig:3}(e) exhibits hole-like Dirac cones along the AFM direction and a hole pocket at $\Gamma$ of mainly $d_{xz}$ character resulting from the mixing of $\alpha_1$ and $\alpha_2$. Evidently, since the MS state singles out the $\mathbf{Q_1}$ ordering vector, the spectrum becomes $C_2$ symmetric as shown in Fig.~\ref{fig:3}(e,h). Because MS breaks the $d_{xz}/d_{yz}$ degeneracy an associated ferro-orbital ordering $n_{xz}>n_{yz}$ results in a splitting of the bands at the $\Gamma$ point as seen from Fig.~\ref{fig:3}(h). No such splitting takes place in the 2Q states with $M_1=M_2$ (at least in the absence of spin-orbit coupling). 
Finally, as the filling is increased further, large nested areas of $\alpha_1$ and $\gamma$ with the electron pockets appear at the FS as seen from Fig~\ref{fig:3}(c). For the present band, at this filling the FS-nesting is the strongest of the presented cases, and the resulting SDW order parameter and the gaps are the largest. As seen from Fig~\ref{fig:3}(f), $\alpha_1$ and $\beta_i$ become fully gapped, and similarly most of the $\gamma$ pocket except small pieces around $k_x=\pm k_y$ which hybridize with the outermost $\alpha_2$ pocket. The dependence of the band reconstruction on the interaction parameters is found in SI.

\begin{figure}[t]
\begin{center}
\includegraphics[width=\columnwidth]{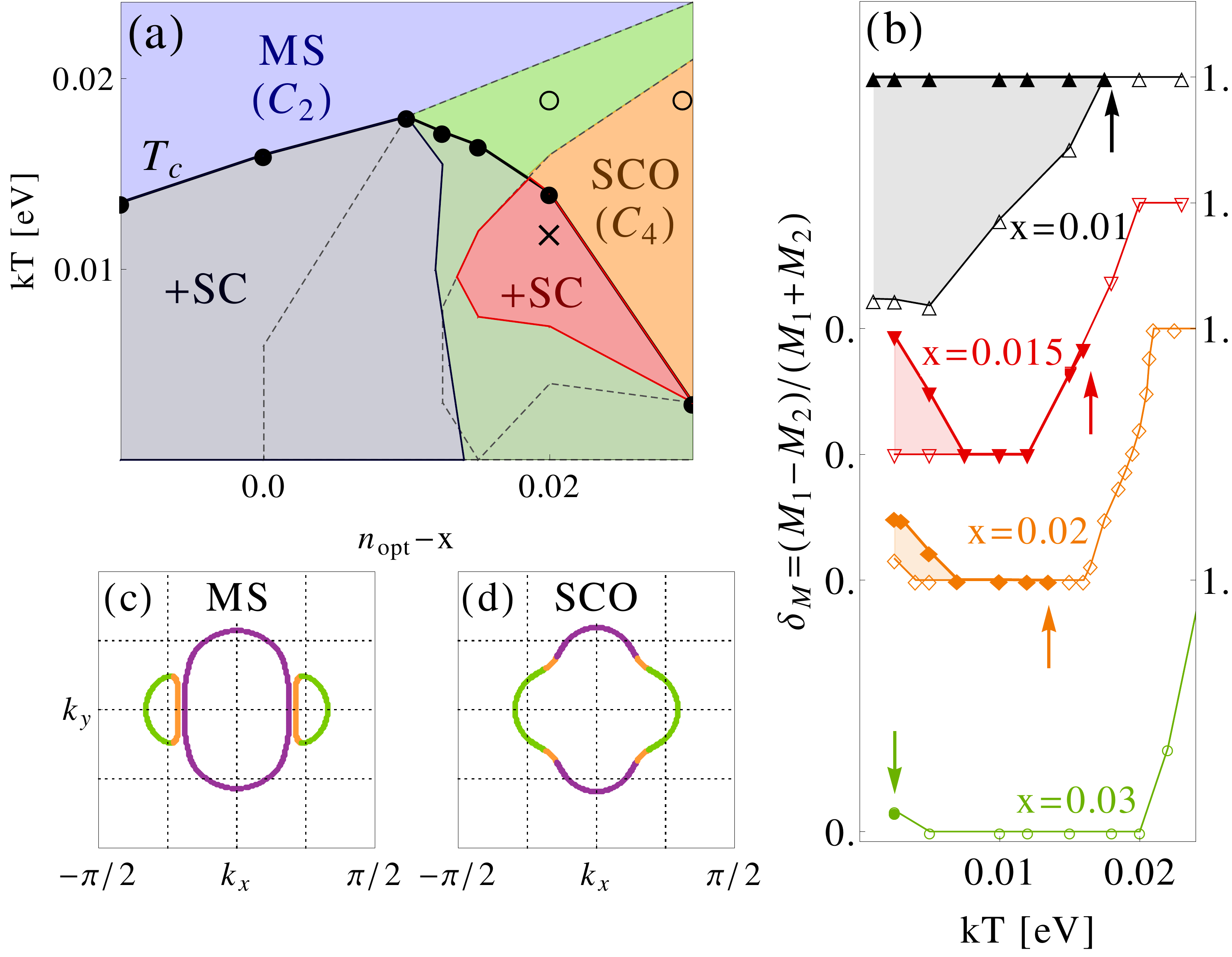}
\end{center}
\caption{(a) Modified phase diagram from the region indicated by the dashed box in Fig.~\ref{fig:1}(e) in the presence of SC order. (b) Magnetic anisotropy $\delta_M=(M_1-M_2)/(M_1+M_2)$ versus $T$ for four different $x$ with (without) SC order shown by the solid (open) symbols. The arrows mark $T_c$. (c,d) Fermi surface at the cross in (a) for MS (c) and SCO (d) order (without SC) using the same orbital color code as in Fig.~\ref{fig:3}.}
\label{fig:4}
\end{figure}

Motivated by the recent experimental discovery of the effects of superconductivity on the magnetic states,\cite{boehmer14} we have included SC order to the model by the following BCS term $\mathcal{H}_{BCS}=\sum_{k,\mu\nu} \Delta_{\mu\nu}(k)\hat{c}_{k\mu\uparrow}^{\dagger} \hat{c}_{-k\nu\downarrow}^{\dagger}$, where $\Delta_{\mu\nu}(k)=\sum_{k',\alpha\beta}\Gamma_{\mu\alpha}^{\beta\nu}(k-k')\langle\hat{c}_{-k'\beta\downarrow}\hat{c}_{k'\alpha\uparrow}\rangle$. A recent Landau order parameter expansion has also been used to study this problem close to $T_N$.\cite{kang14} The effective pairing vertices $\Gamma_{\mu\alpha}^{\beta\nu}(k-k')$ are obtained from the RPA spin and charge susceptibilities in the PM state with leading $s_\pm$ symmetry (see SI for details). 
In order to study the effects of SC on both the 1Q and 2Q magnetic states, we focus on the region of the phase diagram outlined by the dashed box in Fig.~\ref{fig:1}(e), and self-consistently solve the associated Bogoliubov-de Gennes equations including both magnetic and SC order parameters. 
The modified phase diagram shown in Fig.~\ref{fig:4}(a) exhibits a noticeable effect of SC on the boundaries between both magnetic states as seen by comparison to the dashed gray lines indicating the transition lines from Fig.~\ref{fig:1}(e) without SC order. (The phase diagram corresponding to Fig.~\ref{fig:1}(d) including superconducting order is supplied in the SI.) 
Below $T_c$, where both the magnetic and SC order parameters are non-zero, the MS region expands at the expense of the SCO phase. This effect is explicitly shown in Fig.~\ref{fig:4}(b) by the evolution of the magnetic anisotropy $\delta_M=(M_1-M_2)/(M_1+M_2)$, which is a measure of the $C_2$ symmetry breaking, i.e. $\delta_M=0$ ($\delta_M=1$) for the SCO (MS) state, and $\delta_M>0$ for mixed $C_2$ states with $0<M_2<M_1$. As seen from Fig.~\ref{fig:4}(b), without SC $\delta_M$ gradually evolves from $\delta_M=1$ at high $T$ to $\delta_M=0$ at low $T$, with a transition that sharpens with increasing $x$. In the presence of SC order, however, $\delta_M$ is pushed up, as indicated by the shaded regions in Fig.~\ref{fig:4}(b), and the magnetic order is driven towards the MS phase. This effect is particularly pronounced in the regions of large $\Delta_{\mu\nu}(k)$ closer to the MS phase, and is consistent with recent experiments.\cite{boehmer14}

\begin{figure}[b]
\begin{center}
\includegraphics[width=0.9\columnwidth]{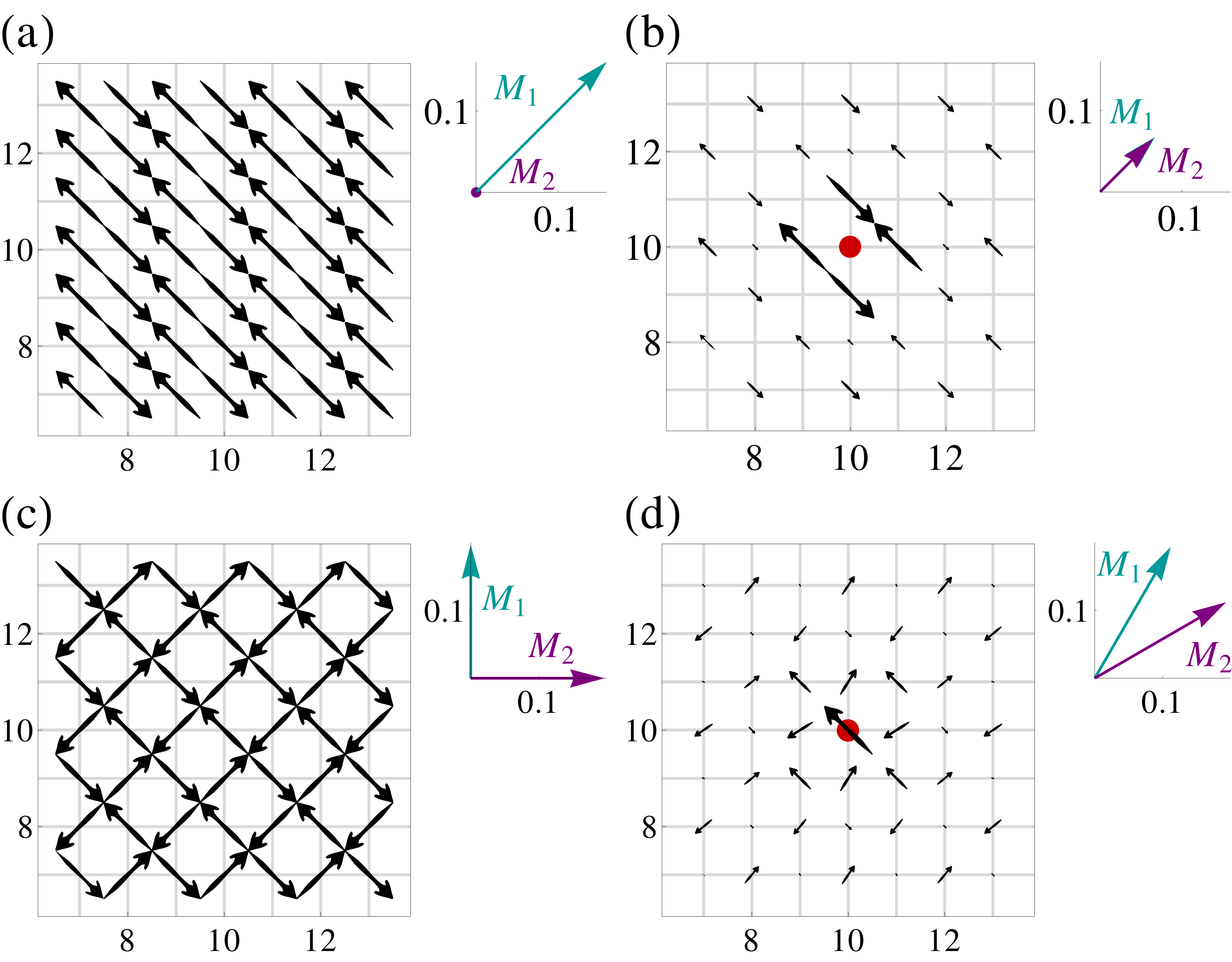}
\end{center}
\caption{(a,c) Magnetization at low $T$ and $x=-0.09$ in the absence of an impurity, and (b,d) in the presence of a single impurity at the center (red dot) with (b) $V=5$ eV, and (d) $V=2$ eV. The associated Fourier components $\mathbf{M_1}$ and $\mathbf{M_2}$ are also shown (the vectors are normalized to (a) $M=0.16$, (b) $M=0.43$, (c) $M=0.2$, and (d) $M=0.68$.)}
\label{fig:5}
\end{figure}

The $T_c$ line shown by the filled black dots in Fig.~\ref{fig:4}(a) evidently exhibits a clear drop across the MS-SCO transition. This reduction of $T_c$ is directly caused by the emergence of the SCO state as verified by the significantly higher $T_c$ (empty circles) found by a separate calculation with the magnetic order forced to the MS type. 
The stronger competition between SC and magnetic order can be explained by a reduced density of states $N(\epsilon)$ at $\epsilon_F$ in the SCO phase (compared to MS), $N_{2Q}(\epsilon_F)\sim0.64 N_{1Q}(\epsilon_F)$. In addition, the dominant SC pairing is the intra-orbital $d_{xy}$ element caused by strong $\gamma$-$\beta_i$ FS nesting (see SI). However, as seen explicitly from Figs.~\ref{fig:4}(d) the FS in the SCO phase (without SC) contains significantly less $d_{xy}$ orbital character (orange points) compared to the corresponding MS FS shown in Fig.~\ref{fig:4}(c). 
In summary $T_c$ is reduced from the PM state into the magnetic phase ($N_{1Q}(\epsilon_F)\sim0.42 N_{PM}(\epsilon_F)$ and $N_{2Q}(\epsilon_F)\sim0.26 N_{PM}(\epsilon_F)$), but enhanced from the SCO phase into the MS phase at lower $T$ in agreement with the experimental finding of Ref.~\onlinecite{boehmer14}. 

We end with a brief study of disorder effects on the three magnetic states (without SC) by including the term $\mathcal{H}_{imp}= \sum_{\mu\sigma}V \hat{c}_{i^*\mu\sigma}^{\dagger}\hat{c}_{i^*\mu\sigma}$ in Eq.~\eqref{eq:H}, relevant for a nonmagnetic impurity at site $i^*$.
Figure~\ref{fig:5} summarizes our main finding: impurities enhance the 2Q phase by nucleating local SCO islands. The left column shows the homogeneous magnetic phase without the disorder potential (MS and OM), and the right column the resulting SDW order in the presence of the impurity. As seen, the charge potential acts like a seed for the charge order associated with the SCO phase, and pushes the magnetic order towards the collinear magnetic SCO structure. This can be more clearly seen from the Fourier components $\mathbf{M_1}$ and $\mathbf{M_2}$ also depicted in Fig.~\ref{fig:5}.

In summary, we have presented a detailed microscopic study of competing magnetic phases in iron pnictides. We have mapped out the phase diagram both in the absence and presence of SC, and found qualitative agreement with recent experimental findings. 

We acknowledge useful discussions with A. V. Chubukov, I. Eremin, R. M. Fernandes, J. Lorenzana, and financial support from a Lundbeckfond fellowship (grant A9318). 

% ===================================================================================== %

% ===================================================================================== %

% ===================================================================================== %

\pagebreak
\widetext
\begin{center}
\textbf{\large Supplemental Materials: "Competing magnetic double-Q phases and superconductivity-induced re-entrance of $C_2$ magnetic stripe order in iron pnictides"}
\end{center}
%%%%%%%%%% Merge with supplemental materials %%%%%%%%%%
%%%%%%%%%% Prefix a "S" to all equations, figures, tables and reset the counter %%%%%%%%%%
\setcounter{equation}{0}
\setcounter{figure}{0}
\setcounter{table}{0}
\setcounter{page}{1}
\makeatletter
\renewcommand{\theequation}{S\arabic{equation}}
\renewcommand{\thefigure}{S\arabic{figure}}
\renewcommand{\bibnumfmt}[1]{[S#1]}
\renewcommand{\citenumfont}[1]{S#1}

Here, we provide the details of the tight-binding band and the mean-field decoupled Hamiltonian both in real and momentum space. 
We show the orbitally resolved order parameters in the single-Q and double-Q phases. In addition, we show the dependence of the band and the Fermi surface reconstruction on the interacting parameters. Finally, we provide the details of the RPA-generated couplings used to generate the superconducting order.

% ------------------------------------------------------------------------------------------------------------------------------------------------------ %
% ===================================================================================== %
\section{model}
\begin{figure}[b]
\begin{center}
\includegraphics[width=0.6\columnwidth]{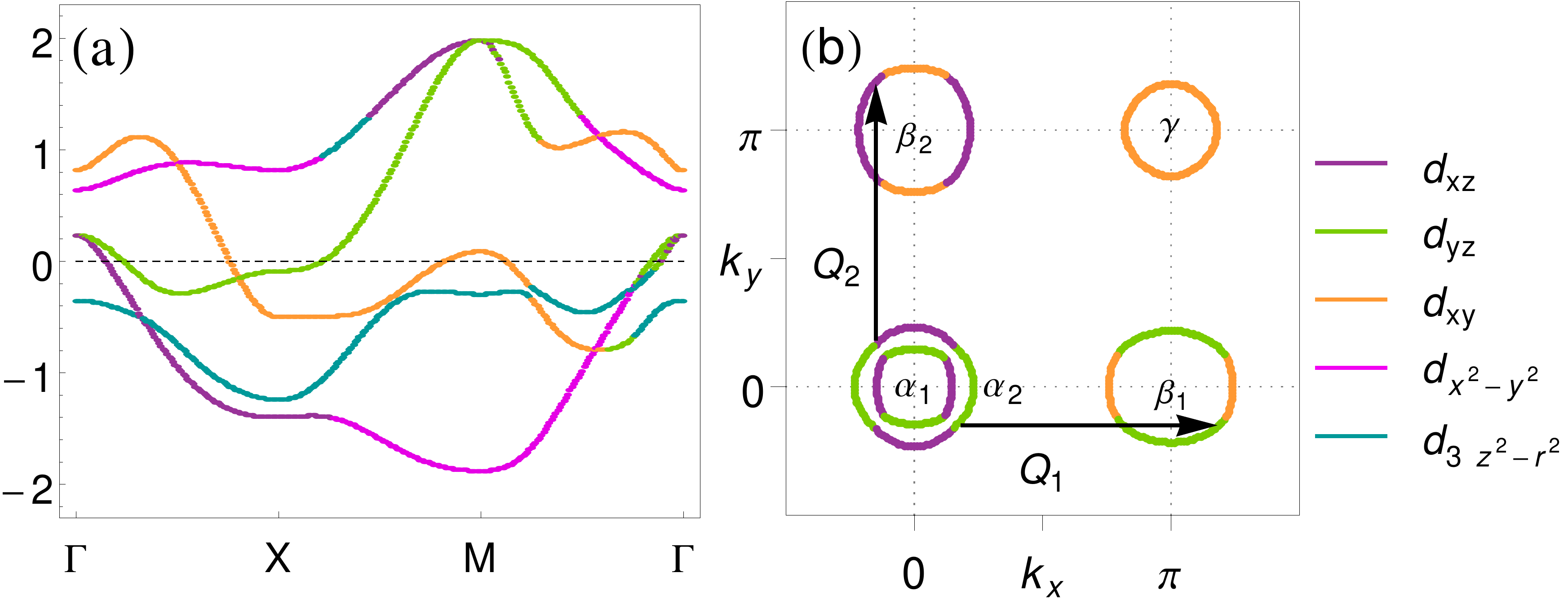}
\end{center}
\caption{(a) Band structure along the high symmetry directions ($\Gamma=(0,0)$, $X=(\pi,0)$, and $M=(\pi,\pi)$), and (b) Fermi surface with main orbital character for the undoped system. }
\label{fig:S1}
\end{figure}

% ------------------------------------------------------------------------------------------------------------------------------------------------------ %
The starting Hamiltonian consists of a five-orbital tight-binding band relevant to the pnictides~\cite{Sikeda10},
\begin{equation}
\label{Seq:H0}
\mathcal{H}_{0}=\sum_{ij,\mu\nu,\sigma}t_{ij}^{\mu\nu}\hat{c}_{i\mu\sigma}^{\dagger}\hat{c}_{j\nu\sigma}-\mu_{0}\sum_{i\mu\sigma}\hat{c}_{i\mu\sigma}^{\dagger}\hat{c}_{i\mu\sigma}.
\end{equation}
The operators $c_{\mathbf{i} \mu\sigma}^{\dagger}$ ($c_{\mathbf{i}\mu\sigma}$) create (annihilate) an electron at site $i$ in orbital state $\mu$ with spin $\sigma$, and $\mu_0$ is the chemical potential which adjusts the filling. 
The indices $\mu$ and $\nu$ denote  the five iron orbitals $d_{xz}$, $d_{yz}$, $d_{xy}$, $d_{x^2-y^2}$, and $d_{z^2}$. The five orbital tight-binding band and Fermi surface for filling $x=-0.09$~\cite{Sikeda10} is shown in Fig.~\ref{fig:S1}. Here, $x$ refers to a filling of $\langle n \rangle = n_{opt} -x$ where $n_{opt}=5.91$ is the optimal doping for magnetism for this band. Thus $x=-0.09$ corresponds to the undoped case with $\langle n \rangle =6.0$.

The interacting part of the Hamiltonian is described by the multi-orbital onsite Hubbard model
\begin{align}
 \label{Seq:int}
 \mathcal{H}_{int}&=U\sum_{i\mu}\hat{n}_{i\mu\uparrow}\hat{n}_{i\mu\downarrow}+U'\sum_{i,\mu<\nu,\sigma}\hat{n}_{i\mu\sigma}\hat{n}_{i\nu\overline{\sigma}}+
(U'-J)\sum_{i,\mu<\nu,\sigma}\hat{n}_{i\mu\sigma}\hat{n}_{i\nu\sigma}\\\nonumber
&\quad+J\sum_{i,\mu<\nu,\sigma}\hat{c}_{i\mu\sigma}^{\dagger}\hat{c}_{i\nu\overline{\sigma}}^{\dagger}\hat{c}_{i\mu\overline{\sigma}}\hat{c}_{i\nu\sigma}
+J'\sum_{i,\mu<\nu,\sigma}\hat{c}_{i\mu\sigma}^{\dagger}\hat{c}_{i\mu\overline{\sigma}}^{\dagger}\hat{c}_{i\nu\overline{\sigma}}\hat{c}_{i\nu\sigma},
\end{align}
with $U'=U-2J$, $J'=J$, and $J=U/4$.

We mean-field decouple Eq.~(\ref{Seq:int}) for all fields $\langle \hat{c}_{i\mu\sigma}^{\dagger} \hat{c}_{j\nu\sigma'}\rangle$ which leads to the following mean-field Hamiltonian
\begin{eqnarray}
\mathcal{H}^{MF}=\sum_{ij\mu\nu}
\begin{pmatrix}
 \hat{c}_{i\mu\uparrow}^{\dagger} & \hat{c}_{i\mu\downarrow}^{\dagger}
\end{pmatrix}
\begin{pmatrix}
\varphi_{ij\uparrow}^{\mu\nu} & \omega_{ii\uparrow}^{\mu\nu}  \\
\omega_{ii\downarrow}^{\mu\nu} & \varphi_{ij\downarrow}^{\mu\nu}  
\end{pmatrix}
\begin{pmatrix}
 \hat{c}_{j\nu\uparrow}\\\hat{c}_{j\nu\downarrow}
\end{pmatrix},
\label{Seq:H}
\end{eqnarray}
where 
\begin{align}
  \varphi_{ij\sigma}^{\mu\nu}&=t_{ij}^{\mu\nu}+\delta_{\mu\nu}[-\mu_{0}+U\langle\hat{n}_{i\mu\overline{\sigma}}\rangle+U'\langle\hat{n}_{i\nu\overline{\sigma}}\rangle+(U'-J) \langle\hat{n}_{i\nu\sigma}\rangle]-(U'-J)\langle \hat{c}_{i\nu\sigma}^{\dagger}\hat{c}_{i\mu\sigma} \rangle +J \langle \hat{c}_{i\nu\overline{\sigma}}^{\dagger}\hat{c}_{i\mu\overline{\sigma}} \rangle+J'\langle \hat{c}_{i\mu\overline{\sigma}}^{\dagger}\hat{c}_{i\nu\overline{\sigma}} \rangle, \\
   \omega_{ii\sigma}^{\mu\nu}&=\delta_{\mu\nu}[-U \langle \hat{c}_{i\mu\overline{\sigma}}^{\dagger}\hat{c}_{i\mu\sigma} \rangle - J\langle \hat{c}_{i\nu\overline{\sigma}}^{\dagger}\hat{c}_{i\nu\sigma} \rangle] -U'\langle \hat{c}_{i\nu\overline{\sigma}}^{\dagger}\hat{c}_{i\mu\sigma} \rangle-J'\langle \hat{c}_{i\mu\overline{\sigma}}^{\dagger}\hat{c}_{i\nu\sigma} \rangle.
\end{align}
Eq.~\eqref{Seq:H} is diagonalized by a unitary transformation  
$\hat{c}_{i\mu\uparrow}= \sum_n u_{i\mu}^n \hat{\gamma}_{n}$ and $\hat{c}_{i\mu\downarrow}=\sum_n \bar u_{i\mu}^n \hat{\gamma}_{n}$ and the following unrestricted fields are obtained self-consistently 
 \begin{align}
  \langle \hat{c}_{i\mu\uparrow}^{\dagger} \hat{c}_{j\nu\uparrow}\rangle&=\sum_{n} u_{i\mu}^{n*}u_{j\nu}^{n} f(E_n), \\
  \langle \hat{c}_{i\mu\downarrow}^{\dagger} \hat{c}_{j\nu\downarrow}\rangle&=\sum_{n} \bar u_{i\mu}^{n*}\bar u_{j\nu}^{n} f(E_n), \\
  \langle \hat{c}_{i\mu\uparrow}^{\dagger} \hat{c}_{i\nu\downarrow}\rangle&=\sum_{n} u_{i\mu}^{n*}\bar u_{i\nu}^{n} f(E_n), \\
  \langle \hat{c}_{i\mu\downarrow}^{\dagger} \hat{c}_{i\nu\uparrow}\rangle&=\sum_{n} \bar u_{i\mu}^{n*}u_{i\nu}^{n} f(E_n), 
 \end{align}
for all sites $i,j$ and orbital combinations $\mu,\nu$. Here $E_n$ denote the eigenvalues, and $f$ is the Fermi function. From these fields we obtain the spin and charge configurations of the final solution in real space
\begin{align}
 M^x(\mathbf{r})&=\sum_{\mu} \left( \langle \hat{c}_{i\mu\uparrow}^{\dagger} \hat{c}_{i\mu\downarrow}\rangle + \langle \hat{c}_{i\mu\downarrow}^{\dagger} \hat{c}_{i\mu\uparrow}\rangle\right),\\
 M^z(\mathbf{r})&=\sum_{\mu} \left(\langle \hat{c}_{i\mu\uparrow}^{\dagger} \hat{c}_{i\mu\uparrow}\rangle-\langle \hat{c}_{i\mu\downarrow}^{\dagger} \hat{c}_{i\mu\downarrow}\rangle\right),\\
 n(\mathbf{r})&=\sum_{\mu} \left(\langle \hat{c}_{i\mu\uparrow}^{\dagger} \hat{c}_{i\mu\uparrow}\rangle+\langle \hat{c}_{i\mu\downarrow}^{\dagger} \hat{c}_{i\mu\downarrow}\rangle\right).
\end{align} 

In order to readily study the electronic properties we also solve the above model in momentum space with the mean-fields $\langle \hat{c}_{\mu\sigma}^{\dagger}(k)\hat{c}_{\nu\sigma'}(k+q_l)\rangle$, where $q_l=\{0,Q_1,Q_2,Q_1+Q_2\}\equiv\{q_0,q_1,q_2,q_3\}$. 
The mean-field Hamiltonian in momentum space takes the following form

\begin{eqnarray}
\sum_{k\mu\neq\nu\sigma}'\Psi^{\dagger}
\begin{pmatrix}
 \xi^{\mu\nu}(k)& W^{\mu\nu}_1 & W^{\mu\nu}_2 &N^{\mu\nu}_3 &\tilde N^{\mu\nu}_0 &\tilde W^{\mu\nu}_1 &\tilde W^{\mu\nu}_2 &\tilde N^{\mu\nu}_3 \\
 +N^{\mu\nu}_0& && & & & & \\
 &\xi^{\mu\nu}(k+q_1) &N^{\mu\nu}_3 &W^{\mu\nu}_{2}&\tilde W^{\mu\nu}_{1} & \tilde N^{\mu\nu}_0&\tilde N^{\mu\nu}_3 &\tilde W^{\mu\nu}_2 \\
 &+N^{\mu\nu}_0 & & & & & & \\
 & & \xi^{\mu\nu}(k+q_2)&W^{\mu\nu}_{1}  & \tilde W^{\mu\nu}_{2}&\tilde N^{\mu\nu}_3 &\tilde N^{\mu\nu}_0 &\tilde W^{\mu\nu}_1 \\
 & &+N^{\mu\nu}_0 & & & & & \\
 & & &\xi^{\mu\nu}(k+q_3)&\tilde N^{\mu\nu}_3 & \tilde W^{\mu\nu}_{2}& \tilde W^{\mu\nu}_{1}&\tilde N^{\mu\nu}_0 \\
 & & &+N^{\mu\nu}_0 & & & & \\
 & & & & \xi^{\mu\nu}(k)& -W^{\mu\nu}_1&-W^{\mu\nu}_2 &N^{\mu\nu}_3 \\
 & & & &+N^{\mu\nu}_0 & & & \\
 & & & & & \xi^{\mu\nu}(k+q_1)& N^{\mu\nu}_3&-W^{\mu\nu}_{2} \\
 & & & & &+N^{\mu\nu}_0 & & \\
 & & H.c. & & & &\xi^{\mu\nu}(k+q_2)&-W^{\mu\nu}_{1} \\
 & & & & & &+N^{\mu\nu}_0 & \\
 & & & & & & &\xi^{\mu\nu}(k+q_3)\\ 
 & & & & & & &+N^{\mu\nu}_0 \\
\end{pmatrix}
\Psi,
\end{eqnarray}
where
\begin{eqnarray}
\Psi^{\dagger}=
 \begin{pmatrix}
  \hat{c}_{\mu\uparrow}^{\dagger}(k) & \hat{c}_{\mu\uparrow}^{\dagger}(k+q_1)&\hat{c}_{\mu\uparrow}^{\dagger}(k+q_2)&\hat{c}_{\mu\uparrow}^{\dagger}(k+q_3)&\hat{c}_{\mu\downarrow}^{\dagger}(k)&\hat{c}_{\mu\downarrow}^{\dagger}(k+q_1)&\hat{c}_{\mu\downarrow}^{\dagger}(k+q_2)&\hat{c}_{\mu\downarrow}^{\dagger}(k+q_3)
 \end{pmatrix},
\end{eqnarray}
and the summation $\sum_k'$ is done in the reduced Brillouin zone $-\pi/2<k_x,k_y<\pi/2$.

\begin{figure}[t]
\begin{center}
\includegraphics[width=0.6\columnwidth]{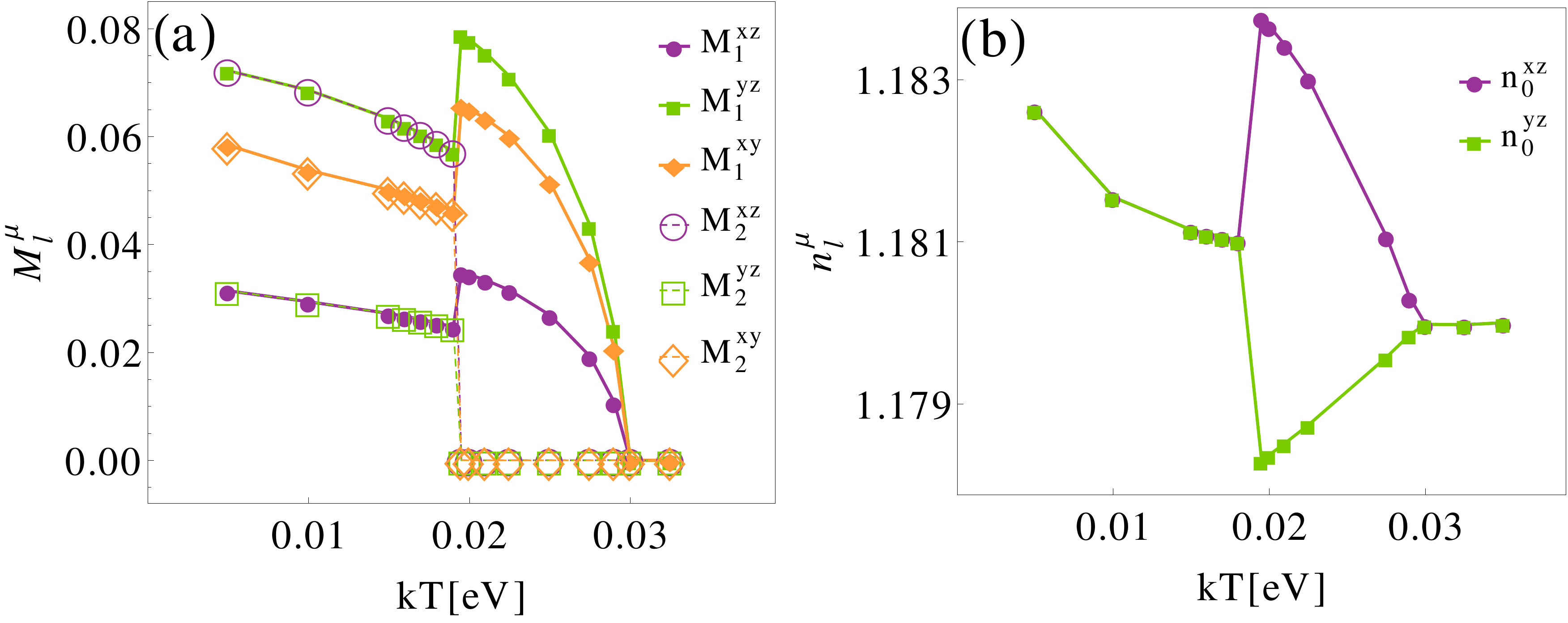}
\end{center}
\caption{Orbitally resolved (a) magnetic OPs for $d_{xz}$, $d_{yz}$ and $d_{xy}$, and (b) charge OPs for $d_{xz}$ and $d_{yz}$ as a function of temperature for $x=-0.09$.}
\label{fig:S2}
\end{figure}

\begin{figure}[t]
\begin{center}
\includegraphics[width=0.94\columnwidth]{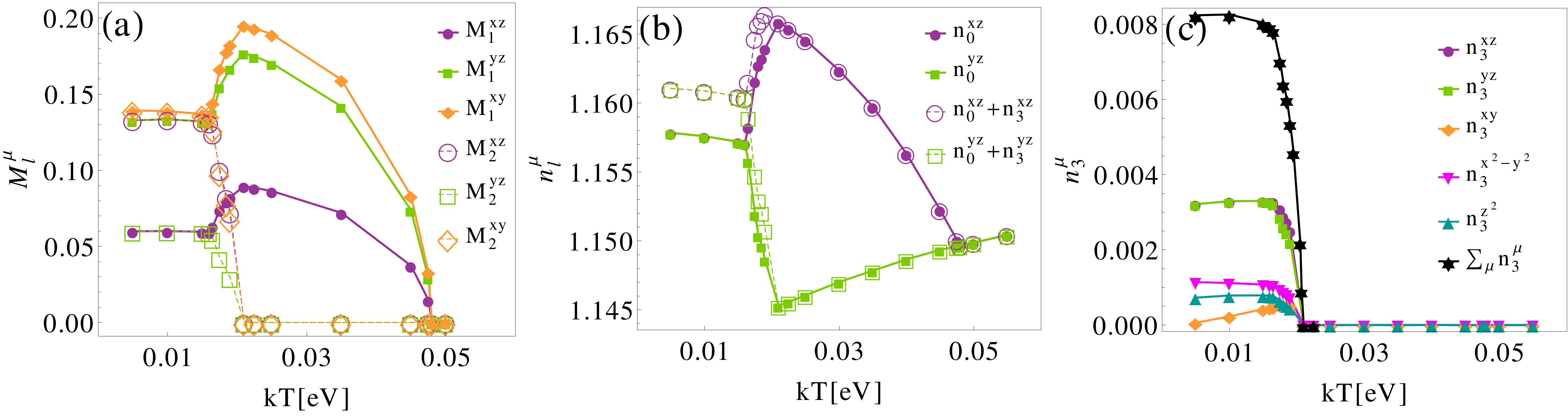}
\end{center}
\caption{Orbitally resolved (a) magnetic OPs for $d_{xz}$, $d_{yz}$ and $d_{xy}$, (b) charge OPs for $d_{xz}$ and $d_{yz}$, and (c) $(\pi,\pi)$ charge OPs for all orbitals as a function of temperature for $x=0.02$. }
\label{fig:S3}
\end{figure}

The entries below the main diagonal were not included for clarity but are obtained by the transpose conjugate of the upper triangular matrix. 
All entries of the mean-field Hamiltonian are defined below
\begin{align}
 \xi^{\mu\nu}(k+q_l)&=\epsilon^{\mu\nu}(k+q_l)-\mu_0\delta_{\mu\nu},\\
 N^{\mu\nu}_0&=\delta_{\mu\nu}\left[ U n^{\mu}_0+(2U'-J)n^{\nu}_0\right]+(-U'+2J)n^{\mu\nu}_0+J'n^{\nu\mu}_0,\\
 W^{\mu\nu}_l&=\delta_{\mu\nu}\left(-U M^{\mu}_l-JM^{\nu}_l \right)-U'M^{\mu\nu}_l-J'M^{\nu\mu}_l,\\
 N^{\mu\nu}_3&=\delta_{\mu\nu}\left[ U n^{\mu}_3+(2U'-J)n^{\nu}_3\right]+(-U'+2J)n^{\mu\nu}_3+J'n^{\nu\mu}_3,\\
 \tilde N^{\mu\nu}_0&=\delta_{\mu\nu}\left(-U\tilde{n}^{\mu}_0-J\tilde{n}^{\nu}_0\right)-U'\tilde{n}^{\mu\nu}_0-J'\tilde{n}^{\nu\mu}_0,\\
 \tilde W^{\mu\nu}_l&=\delta_{\mu\nu}\left(-U\tilde{M}^{\mu}_l-J\tilde{M}^{\nu}_l\right)-U'\tilde{M}^{\mu\nu}_l-J'\tilde{M}^{\nu\mu}_l,\\
 \tilde N^{\mu\nu}_3&=\delta_{\mu\nu}\left(-U\tilde{n}^{\mu}_3-J\tilde{n}^{\nu}_3\right)-U'\tilde{n}^{\mu\nu}_3-J'\tilde{n}^{\nu\mu}_3.
\end{align}
Using the unitary transformation $ \hat{c}_{\mu\uparrow}(k+q_l)=\sum_n u_{l\mu}^n(k)\gamma_n $ and $\hat{c}_{\mu\downarrow}(k+q_l)=\sum_n \bar u_{l\mu}^n(k)\gamma_n$ (where $l=0,1,2,3$) the mean-fields are then self-consistently obtained from the relations 
\begin{align}\nonumber
 n^{\mu\nu}_0&=\sum_{k\sigma}\langle\hat{c}_{\mu\sigma}^{\dagger}(k) \hat{c}_{\nu\sigma}(k)\rangle=\sum_{kn}'\sum_{l=0}^3\left[u_{l\mu}^{n*}(k)u_{l\nu}^{n}(k)+\bar u_{l\mu}^{n*}(k)\bar u_{l\nu}^{n}(k)\right] f_n,\\\nonumber
 n^{\mu\nu}_3&=\sum_{k\sigma}\langle\hat{c}_{\mu\sigma}^{\dagger}(k) \hat{c}_{\nu\sigma}(k+q_3)\rangle=\sum_{kn}'\left\{\sum_{l=1,4}\left[u_{l\mu}^{n*}(k)u_{\bar l\nu}^{n}(k)+\bar u_{l\mu}^{n*}(k)\bar u_{\bar l\nu}^{n}(k)\right]+\sum_{l=2,3}\left[u_{l\mu}^{n*}(k)u_{\bar l\nu}^{n}(k)+\bar u_{l\mu}^{n*}(k)\bar u_{\bar l\nu}^{n}(k)\right] \right\}f_n,\\\nonumber
 M^{\mu\nu}_{1}&=\sum_{k\sigma}\sigma\langle\hat{c}_{\mu\sigma}^{\dagger}(k) \hat{c}_{\nu\sigma}(k+q_{1})\rangle=\sum_{kn}'\left\{\sum_{l=1,2}\left[u_{l\mu}^{n*}(k)u_{\bar l\nu}^{n}(k)-\bar u_{l\mu}^{n*}(k)\bar u_{\bar l\nu}^{n}(k)\right]+\sum_{l=3,4}\left[u_{l\mu}^{n*}(k)u_{\bar l\nu}^{n}(k)-\bar u_{l\mu}^{n*}(k)\bar u_{\bar l\nu}^{n}(k)\right] \right\}f_n,\\\nonumber
M^{\mu\nu}_{2}&=\sum_{k\sigma}\sigma\langle\hat{c}_{\mu\sigma}^{\dagger}(k) \hat{c}_{\nu\sigma}(k+q_{2})\rangle=\sum_{kn}'\left\{\sum_{l=1,3}\left[u_{l\mu}^{n*}(k)u_{\bar l\nu}^{n}(k)-\bar u_{l\mu}^{n*}(k)\bar u_{\bar l\nu}^{n}(k)\right]+\sum_{l=2,4}\left[u_{l\mu}^{n*}(k)u_{\bar l\nu}^{n}(k)-\bar u_{l\mu}^{n*}(k)\bar u_{\bar l\nu}^{n}(k)\right] \right\}f_n,\\\nonumber
 \tilde n^{\mu\nu}_0&=\sum_{k\sigma}\sigma\langle\hat{c}_{\mu\bar\sigma}^{\dagger}(k) \hat{c}_{\nu\sigma}(k)\rangle=\sum_{kn}'\sum_{l=0}^3\left[ \bar u_{l\mu}^{n*}(k) u_{l\nu}^{n}(k)-u_{l\mu}^{n*}(k) \bar u_{l\nu}^{n}(k)\right]f_n,\\\nonumber
 \tilde n^{\mu\nu}_3&=\sum_{k\sigma}\sigma\langle\hat{c}_{\mu\bar\sigma}^{\dagger}(k) \hat{c}_{\nu\sigma}(k+q_3)\rangle=\sum_{kn}'\left\{\sum_{l=1,4}\left[\bar u_{l\mu}^{n*}(k)u_{\bar l\nu}^{n}(k)-\bar u_{l\mu}^{n*}(k) u_{\bar l\nu}^{n}(k)\right]+\sum_{l=2,3}\left[\bar u_{l\mu}^{n*}(k)u_{\bar l\nu}^{n}(k)+ u_{l\mu}^{n*}(k)\bar u_{\bar l\nu}^{n}(k)\right] \right\}f_n,\\\nonumber
 \tilde M^{\mu\nu}_{1}&=\sum_{k\sigma}\langle\hat{c}_{\mu\bar\sigma}^{\dagger}(k) \hat{c}_{\nu\sigma}(k+q_{1})\rangle=\sum_{kn}'\left\{\sum_{l=1,2}\left[\bar u_{l\mu}^{n*}(k)u_{\bar l\nu}^{n}(k)+ u_{l\mu}^{n*}(k)\bar u_{\bar l\nu}^{n}(k)\right]+\sum_{l=3,4}\left[\bar u_{l\mu}^{n*}(k)u_{\bar l\nu}^{n}(k)+ u_{l\mu}^{n*}(k)\bar u_{\bar l\nu}^{n}(k)\right] \right\}f_n,\\\nonumber
 \tilde M^{\mu\nu}_{2}&=\sum_{k\sigma}\langle\hat{c}_{\mu\bar\sigma}^{\dagger}(k) \hat{c}_{\nu\sigma}(k+q_{2})\rangle=\sum_{kn}'\left\{\sum_{l=1,3}\left[\bar u_{l\mu}^{n*}(k)u_{\bar l\nu}^{n}(k)+ u_{l\mu}^{n*}(k)\bar u_{\bar l\nu}^{n}(k)\right]+\sum_{l=2,4}\left[\bar u_{l\mu}^{n*}(k)u_{\bar l\nu}^{n}(k)+ u_{l\mu}^{n*}(k)\bar u_{\bar l\nu}^{n}(k)\right] \right\}f_n,
 \end{align}
where the abbreviation $f_n\equiv f(E_n(k))$ has been used.

\section{Orbitally resolved order parameters: $M_l^{\mu}$ and $n_l^{\mu}$}

The Figs.~\ref{fig:S2} and \ref{fig:S3} show the orbital content of the magnetic and charge order parameters versus $T$ corresponding to the results presented in Fig.~2 of the main text. The largest magnetic order parameter (OP) components $M_l^{\mu}$ correspond to the best nested orbitals, $d_{xz}/d_{yz}$ for $x=-0.09$ in Fig.~\ref{fig:S2}(a), and $d_{xy}$ for the higher filling of $x=0.02$ in Fig.~\ref{fig:S3}(a) where the $\gamma-\beta_i$ nesting has improved.
The remaining orbital components are not shown for presentational simplicity. The ferro-orbital order ($n_0^{xz} > n_0^{yz}$) can be seen across the PM-MS transition in Figs.~\ref{fig:S2}(b) and \ref{fig:S3}(b), and it collapses upon formation of both OM and SCO order. In addition, a small $q_3\equiv \mathbf{Q_1}+\mathbf{Q_2}=(\pi,\pi)$ charge order develops as soon as $M_2>0$ in the SCO phase. 
Figure~\ref{fig:S3}(c) shows the orbital character of this charge order, and the dominant contributions arise from the $d_{xz}$ and $d_{yz}$ orbitals.

\section{Band reconstruction and interaction parameters}

\begin{figure}
\begin{center}
\includegraphics[width=0.65\columnwidth]{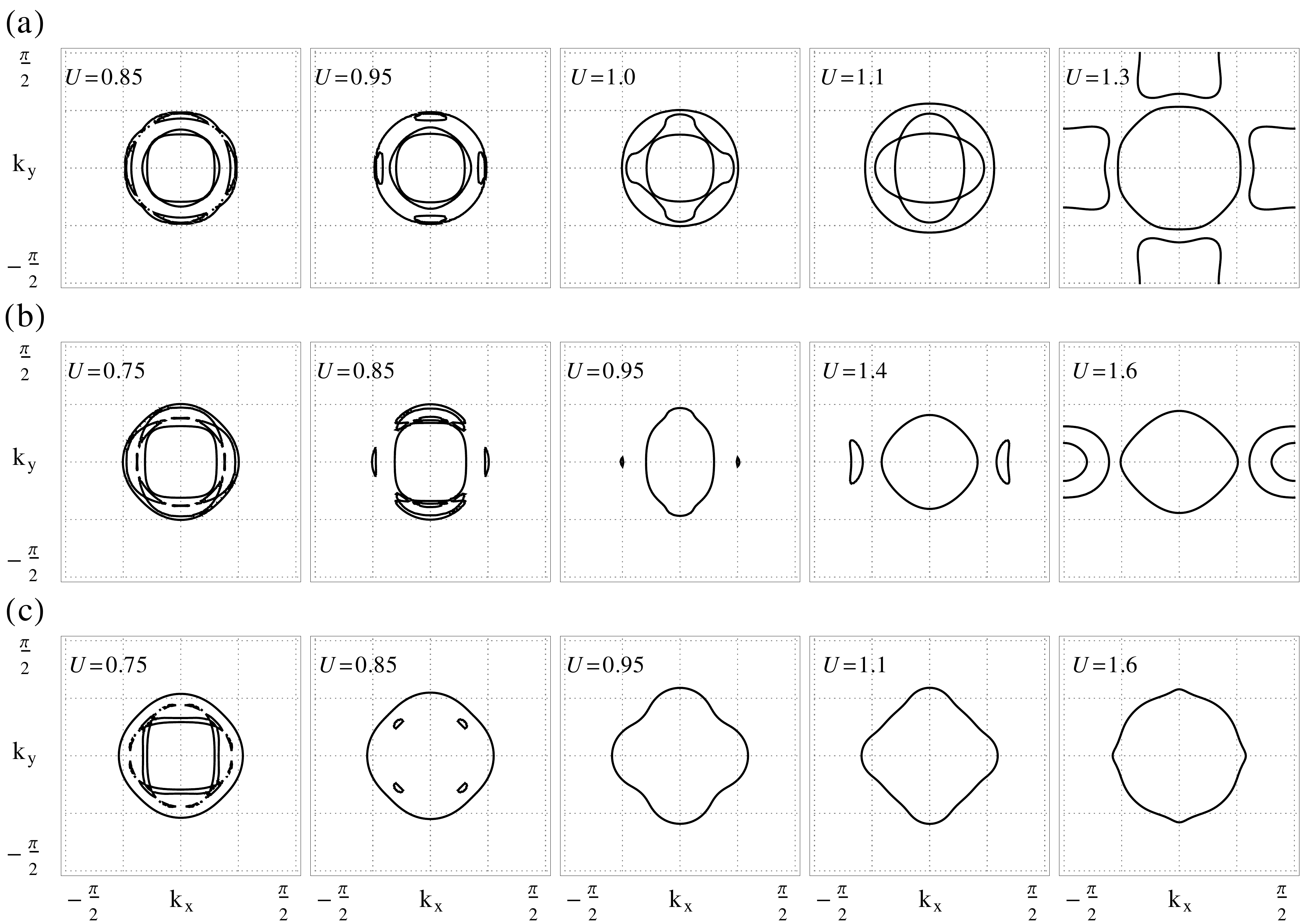}
\end{center}
\caption{Fermi surfaces for different values of the interacting parameter $U$ in the (a) $x=-0.09$ and OM, (b) $x=-0.04$ and MS, and (c) $x=0.02$ and SCO. }
\label{fig:S4}
\end{figure}

The band reconstruction taking place in the ordered state obviously depends on the amplitude of the magnetic OPs $M_1$ and $M_2$, which in turn depend on the interacting strength $U$. 
This is illustrated in Fig.~\ref{fig:S4} where the reconstructed FSs in the (a) OM, (b) MS and (c) SCO states are plotted as a function of $U$. For comparison the paramagnetic FS is shown in the first panel of each case. As seen, the resulting reconstructed Fermi surface depends significantly on $U$. 

\section{Superconducting pairing vertex}

\begin{figure}
\begin{center}
\includegraphics[width=0.75\columnwidth]{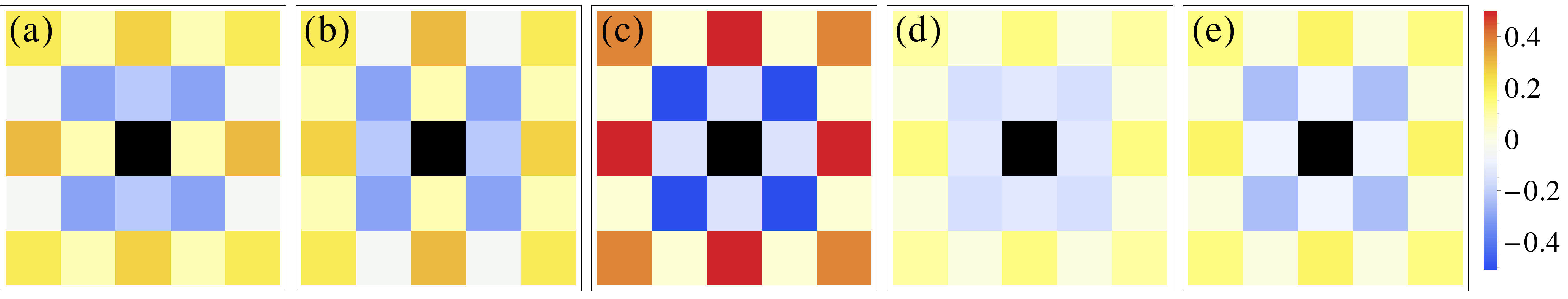}
\end{center}
\caption{Spatial dependence of the intra-orbital effective pairing constants $\Gamma^{\mu\mu}_{\mu\mu}(\mathbf{r_{ij}})$ in eV from the central site. (a) $d_{xz}$ (b) $d_{xz}$, (c) $d_{xy}$, (d) $d_{x^2-y^2}$ and (e) $d_{z^2}$.}
\label{fig:S5}
\end{figure}

The multi-orbital pairing vertex in the singlet channel~\cite{graser:2009} is calculated from the RPA spin- $\chi^{RPA}_s$ and charge-susceptibilities $\chi^{RPA}_c$,
\begin{align}
\label{Seq:gammak}
 \Gamma^{st}_{pq}(k-k',0)&= \bigg[ \frac{3}{2} U^{s}\chi_{s}^{RPA} (k-k',0) U^{s} + \frac{1}{2} U^{s}-\frac{1}{2} U^{c} \chi_{c}^{RPA}(k-k',0)U^{c} + \frac{1}{2}U^{c} \bigg]_{pq}^{st}
\end{align}
where $U^s$ and $U^c$ are $5\times5$ matrices identical to those of Ref.~\onlinecite{graser:2009}. 
The real-space pairings are then obtained by Fourier transforming equation \eqref{Seq:gammak}, $\Gamma_{\mu\alpha}^{\beta\nu}(\mathbf{r_{ij}})=\sum_{\mathbf{q}} \Gamma_{\mu\alpha}^{\beta\nu}({\mathbf{q}}) \exp(i{\mathbf{q}}\cdot({\mathbf{r_i}}-{\mathbf{r_j}}))$. 
We retain all orbital combinations up to next-nearest neighbor sites to calculate the superconducting order parameter $\Delta_{\mathbf{ij}}^{\mu\nu}=\sum_{\alpha\beta}\Gamma_{\mu\alpha}^{\beta\nu}(\mathbf{r_{ij}})\langle\hat{c}_{\mathbf{j}\beta\downarrow}\hat{c}_{\mathbf{i}\alpha\uparrow}\rangle$.\cite{gastiasoro2013} 
The intra-orbital effective pairings are shown in Fig.~\ref{fig:S5}.

\section{Phase diagram including superconductivity in the low-$U$ limit}

\begin{figure}
\begin{center}
\includegraphics[width=0.5\columnwidth]{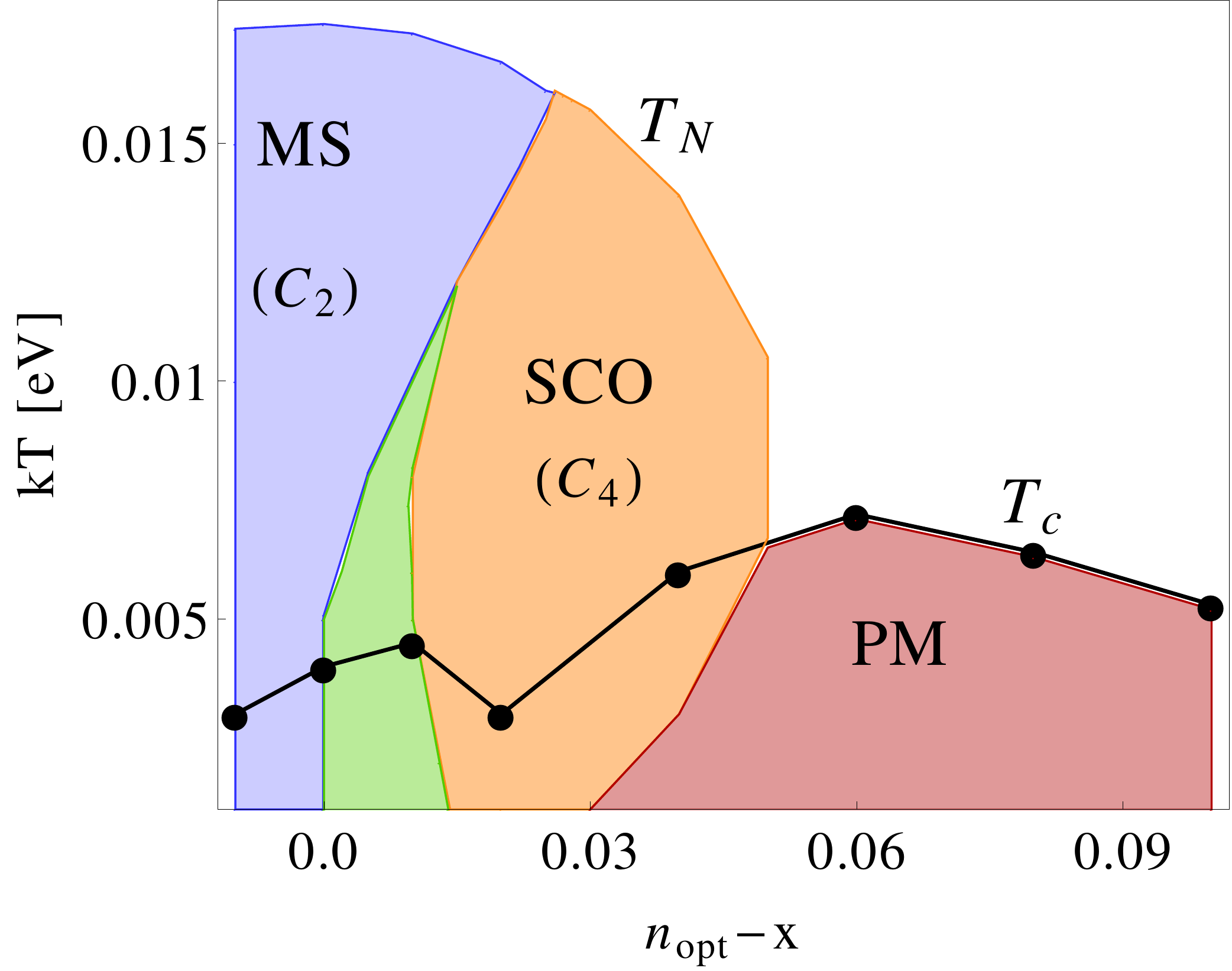}
\end{center}
\caption{Phase diagram showing the magnetic and superconducting phases as a function of $T$ and filling $n_{opt} -x$ for $U = 0.85$ eV. }
\label{fig:S6}
\end{figure}

In addition to the phase diagrams shown in the main text, we have also calculated the phase diagram in the presence of competing superconductivity at $U=0.85$ eV shown in Fig.~\ref{fig:S6}. The corresponding normal state phase diagram for this case is shown in Fig. 1(d) of the main text. The lower value of $U$ pushes the magnetic structure to lower values of the doping which prevents the occurrence of the IC magnetic phase, and the paramagnetic superconducting phase directly merges with the SCO $C_4$-magnetic phase in this case. Note that for the particular parameters used to generate Fig.~\ref{fig:S6} there is no superconductivity-induced re-entrance of the $C_2$ phase, which we attribute to the pairing being too weak to cause a switch of the preferred magnetic structure.

% ===================================================================================== %

\end{document}